\documentclass[%
 reprint, 
 amsmath,amssymb,
 aps, 
pra,
]{revtex4-2}

\usepackage{graphicx}% Include figure files
\usepackage{dcolumn}% Align table columns on decimal point
\usepackage{bm}% bold math

\usepackage{siunitx}
\DeclareSIUnit\Erec{\ensuremath{\textit{E}_\text{rec}}}
\usepackage{graphicx,float}

\usepackage[dvipsnames]{xcolor}
\usepackage{physics}
\usepackage{braket}
\usepackage{float}
\usepackage{csquotes}
\usepackage[english]{babel}
\usepackage[colorlinks=true, citecolor = MidnightBlue, linkcolor = MidnightBlue, urlcolor=MidnightBlue, pdfborder={0 0 0}]{hyperref}
\usepackage{orcidlink}
\usepackage{times}
\usepackage{newtxmath}

\newcommand\df[2]{\displaystyle\frac{#1}{#2}}

\usepackage{dsfont}

\begin{document}

\title{Coherent coupling of momentum states: selectivity and phase control}

%Coherent coupling of momentum states: selectivity and phase control

\author{Charlie Leprince\,\orcidlink{0009-0002-5490-6767}}
 \email{Contact author: charlie.leprince@protonmail.com}
\author{Victor Gondret\,\orcidlink{0009-0005-8468-161X}}
\author{Clothilde Lamirault\,\orcidlink{0009-0001-6468-2181}}
\author{Rui Dias\,\orcidlink{0009-0004-4158-7693}}
\author{Quentin Marolleau\,\orcidlink{0009-0002-3587-3912}}
 \altaffiliation[Present address ]{Qblox, Delftechpark, Netherlands.}%Lines break automatically or can be forced with \\
\author{Denis Boiron\,\orcidlink{0000-0002-2719-5931}}
\author{Christoph I. Westbrook\,\orcidlink{0000-0002-6490-0468}}
\affiliation{%
  Université Paris-Saclay, Institut d’Optique Graduate School, CNRS, Laboratoire Charles Fabry, 91127, Palaiseau, France
}

\begin{abstract}
	\noindent
We demonstrate the effect of pulse shaping in momentum selective atomic Bragg diffraction. 
We compare temporal square pulses, which produce sidelobes in momentum space, with other shapes which can produce more nearly square momentum distributions. 
We produce pulses that simultaneously address two sets of velocity classes and demonstrate that we can control the differential phase imprinted on them in a way that is insensitive to laser phase fluctuations. 
Our work marks a significant step forward in testing Bell inequalities using massive particles entangled in momentum.

\end{abstract}
%\pacs{}
%\maketitle

%%%%%%%%%%%%%%
%   Message
%%%%%%%%%%%%%%
\maketitle

\section{Introduction}
The coherent coupling of quantum states is central to many quantum technologies including quantum computation, simulation and sensing~\cite{degen_quantum_2017, cronin_2009_optics, Saffman_2010}.
Depending on the specific application, this coupling must typically be optimized according to various criteria such as efficiency, selectivity, speed or immunity from noise. Here we will discuss a common example, the coupling of different atomic momentum states using Bragg diffraction 
or momentum selective Raman transitions~\cite{miffre_atom_2006,cronin_2009_optics, Bord2001TheoreticalTF}.

These coupling mechanisms can be understood as two-photon transitions producing transfers between two well defined momentum classes~\cite{martin.1988.bragg}.
In the Raman case, the momentum transfer is accompanied by a transition between two low lying states in a three-level system~\cite{Kasevich_1991}.
They are basic techniques in atom interferometry
\cite{Giltner_1995, Kasevich_1991} as well as being a spectroscopy technique for many body physics~\cite{ozeri.2005.colloquim}.
Laser beams producing the transfer are typically pulsed on for some duration and roughly speaking, the duration determines the momentum selectivity of the pulse.
Bragg diffraction has been used to perform atomic Hong-Ou-Mandel and other interferometry experiments which are working towards a Bell inequality test with momentum entangled atoms~\cite{Lopes_atomic_2015,Dussarrat_two-particle_2017, shin_bell_2019, thomas_matter-wave_2022}. 
In these experiments, both the momentum selection and the control of the wavepacket phase are crucial.
The transition probability associated with a pulse depends on the momentum of the atoms. In particular, the atomic response to a pulse whose temporal profile is square leads to a transfer efficiency which is not flat near the resonant momentum class, and which also contains sidelobes out of resonance (see Eq.~\ref{eq:FourierTransform}).
 
 These drawbacks can be mitigated by choosing more complex pulse shapes.
Some authors have investigated the use of Gaussian pulse shapes~\cite{muller_atom-wave_2008,fang_improving_2018,zhao_optimized_2022} and polychromatic frequency spectra~\cite{lellouch_polychromatic_2023} while others have used optimal control techniques~\cite{alway_arbitrary_2007,saywell_efficient_2022,saywell_enhancing_2023,dedes_optimizing_2023,louie_robust_2023} to improve various aspects of interferometer performance.
While highly effective couplings can be engineered through optimal control techniques, the cost function used for the optimization is very specific and the resulting waveform is complex. 
Pulse shaping protocols were pioneered in the context of NMR~\cite{geen_1991_band-selective,mcdonald_uses_1991} and have recently found  some applications in atom interferometry~\cite{dunning_composite_2014,luo_contrast_2016,wang_amplitude-modulated_2024}.
These earlier methods have the advantages of being general, with analytical forms and depending on a small number of parameters. However, they require changing the sign of the two-photon Rabi frequency during the pulse 
(see Eq.~\ref{eq:sinc}).
%, which is technically challenging. 
%\charlie{Indeed, most experimental setups control the Rabi frequency  through the power of the beams, as its modulus  is proportional to the light intensity. Adding a phase control is more demanding experimentally, because of laser phase fluctuations for instance.}
%\chris{peut-etre ces phrases ne sont pas neccessaires ...}

In this article we report the experimental realization of these pulse shaping techniques in the context of atomic Bragg diffraction.
We first demonstrate our ability to efficiently address atoms in a given momentum class while suppressing the coupling to others.  
We also extend these ideas to implement a simple and effective method to simultaneously address two sets of momentum classes and control their relative phase.

%%%%%%%%%%%%%%%%%%
% Calculations
%%%%%%%%%%%%%%%%%%
\vskip 6pt
%\textbf{Bragg diffraction model}\\

\section{Model and calculations}

Bragg diffraction can be understood as a two-photon transition coupling momenta separated by $2\hbar k$, with $k=\frac{2\pi}{\lambda}\sin\frac{\alpha}{2}$, where $\alpha$ is the angle between the beams, each characterized by a frequency $\omega_i$, a phase $\varphi_i$, and a Rabi frequency $\Omega_i$. 
%\cite{kozuma_coherent_1999}. 
In the rotating frame, two momenta are coupled by an interaction Hamiltonian:
\begin{equation}
\hat{H}_{\mathrm{I}}=\displaystyle\frac{\hbar\Omega_{\mathrm{R}}(t)}{2}\,\mathrm{e}^{\mathrm{i}\delta t}\ket{p}\bra{p+2\hbar k} + h.c.
\label{eq:Braggcoupling}
\end{equation}
where {\it h.c.} denotes the Hermitian conjugate, and $\Omega_{\mathrm{R}}$ is the two-photon Rabi frequency, defined as
\begin{equation}
	\Omega_{\mathrm{R}}=\displaystyle\frac{\Omega_1\Omega_2^{\ast}}{2\Delta}=\displaystyle\frac{|\Omega_1|\cdot|\Omega_2|}{2\Delta}\,\mathrm{e}^{\mathrm{i}\varphi_L}
%\propto I\,\mathrm{e}^{\mathrm{i}\Delta\varphi}
	\label{eq:rabifreq}
\end{equation}
with $\varphi_L= \varphi_1 - \varphi_2$ the laser phase difference. 
We also define
the two-photon detuning $\delta$
\begin{equation}
	\hbar\delta=\hbar(\omega_2-\omega_1)-\left(\dfrac{2\hbar^2 k^2}{m}+\dfrac{2\hbar k}{m}p\right)
\label{eq:detuning_expression}
\end{equation}
which is assumed to be small compared to $\Delta$, the one-photon detuning from the excited state. The doublet is resonantly coupled when the frequency difference $\omega_2-\omega_1$ and the momentum $p$ are such that $\delta=0$. 
%A doublet $(p', p'+2\hbar k)$ with $p'\neq p$ will still interact, 
Off resonant doublets are still coupled, but their transfer efficiencies are lower, a point
%of the off-resonance doublets will differ from the resonance efficiency, 
which is of central importance for this article. %, a particular momentum doublet $(p,p+2\hbar k)$.\\
This Hamiltonian $\hat{H}_{\mathrm{I}}$ can be derived from the dipole atom-light interaction of two beams, after adiabatic elimination of the excited state due to the large detuning $\Delta$~\cite{Giese2013}.

Given an atom in an initial momentum state $\ket p$, the above interaction Hamiltonian takes the atom to the state $c_p\ket p + c_{p+2\hbar k} \ket {p+2\hbar k}$.
Assuming that $\delta$ is constant and that $c_{p+2\hbar k}$ remains small, 1st order perturbation theory predicts
 \begin{equation}
c_{p+2\hbar k}(\delta) \propto \int_0^t   \mathrm{d} t' \, \Omega_{\mathrm{R}}(t') \,\mathrm{e}^{\mathrm{i} \,\delta t' }
\label{eq:FourierTransform}
\end{equation}
meaning that the deflection coefficient $c_{p+2\hbar k}$ as a function of the detuning is proportional to the Fourier transform  of the pulse $\Omega_{\mathrm{R}}$ as a function of time. Thus a square pulse results in a momentum space profile in the form of a sinc function ($\mathrm{sinc}(x)=\sin x / x$). Conversely, one can realize a transfer with a nearly square profile in momentum space by having the atoms interact with a laser pulse whose profile is a sinc function.

When the fraction of transferred atoms $|c_{p+2\hbar k}|^2 $ is large, Eq.~\ref{eq:FourierTransform} is not exact; however we will show that even for a 50\% or 100\% transfer, 
%we can use the above insight to design 
a sinc is a simple and effective pulse shape in our conditions. 
In the following, we will denote $\Omega_{\mathrm{M}}$ as the magnitude of the two-photon Rabi frequency, so that a square pulse corresponds to $\Omega_{\mathrm{R}}(t)=\Omega_{\mathrm{M}}$ during the pulse and $\Omega_{\mathrm{R}}(t)=0$ otherwise.
According to the Fourier relationship (see Eq.~\ref{eq:detuning_expression} and~\ref{eq:FourierTransform}), the selected momentum spectrum contains sidelobes at  momenta inversely proportional to the duration of the pulse.
A temporal sinc pulse  in the interval $[0,T]$ is given by:
\begin{equation}
\Omega_{\mathrm{R}}(t)=\Omega_{\mathrm{M}}\, \mathrm{sinc}(\Omega_{\mathrm{S}}(t-T/2)).
\label{eq:sinc}
\end{equation}
The duration $T$ of the sinc pulse has been chosen relative to the typical $1/\Omega_{\mathrm{S}}$ pulse oscillation period, so as to retain a significant number of sinc sidelobes (at least 3 for the desired momentum response) while ensuring that the pulse is short enough to avoid decoherence issues due to spontaneous emission (see Fig.~\ref{fig:setup}(a) for an example).
In order to produce a deflector (100\% transfer) one chooses $\Omega_{\mathrm{S}}=\Omega_{\mathrm{M}}$ so that the time integral of the Rabi frequency is $\pi$.
For a 50-50 beam splitter one chooses $\Omega_{\mathrm{S}}=2\,\Omega_{\mathrm{M}}$.
%Although Eq.~\ref{eq:FourierTransform} is not exact, 
The use of a sinc pulse has the advantage, compared to optimal control methods,
%\cite{saywell_efficient_2022, saywell_enhancing_2023, alway_arbitrary_2007} 
of being intuitive and having a simple analytical form.  One can also realize more complex pulse shapes, see Eq. \ref{eq:reburp} and Fig. \ref{fig:mirror}.

In the above discussion, it is assumed that there is no diffraction into higher orders, {\it i.e.}, that we remain within the so-called \textit{Bragg} regime~\cite{cronin_2009_optics}. 
This limits the peak power of the beams, so that the peak transfer energy $\hbar \, \Omega_{\mathrm{M}}$ remains below the two photon recoil energy $\hbar^2k^2/m$. 
The pulse parameters are chosen so that this condition, where only two diffraction orders are coupled,  is well satisfied. This is checked experimentally by counting atoms at momentum $p-2\hbar k$, $p+4\hbar k$, and so on.

%When modeling the diffraction process, higher orders can be included. 
For a further confirmation, numerical calculations are conducted using a multi-level model where levels are coupled two by two through the interaction Hamiltonian from Eq.~\ref{eq:Braggcoupling}, {\it i.e.} without making a two-level approximation. 
The full Hamiltonian used for the simulations~\footnote{The numerical results were obtained using the \href{https://docs.scipy.org/doc/scipy/reference/generated/scipy.integrate.odeint.html}{\texttt{integrate.odeint}} solver of the Python SciPy package. The Hamiltonian of Eq.~\ref{eq:hsim} was truncated at $n\in[-2,2]$. Initially, the state is   such that all the atoms are in the considered momentum class $p$. The numerical figures are plotted as a function of $\delta_0/2\pi$, which determines the resonant doublet.} is therefore:
\begin{equation}\label{eq:hsim}\hat{H}_{\mathrm{sim}}=\displaystyle\df{\hbar\Omega_{\mathrm{R}}(t)}{2}\sum_n\mathrm{e}^{\mathrm{i}\delta_{2n}t} \ket{p+2n\hbar k}\bra{p+2(n+1)\hbar k}\end{equation}
where
\begin{equation}\hbar \delta_{2n}=\hbar(\omega_2-\omega_1)-\df{2\hbar k}{m}\left[\hbar k(2n+1)+p\right]
\end{equation}
%\noindent This Hamiltonian was truncated at $n\in[-2,2]$ (boundaries for which the corresponding states are never populated in the range of intensity explored). 

\noindent In the results shown below, the higher order terms have negligible effect.
\section{Experimental setup}

\begin{figure*}[btp]
\begin{center}
\includegraphics[width=1\linewidth]{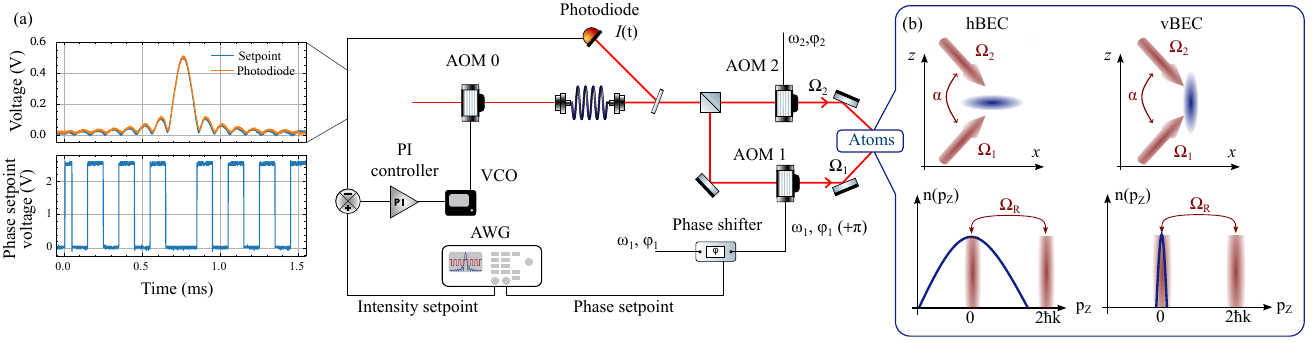}
\caption{(a) Schematic diagram of the modulation technique to produce a sinc-shaped excitation. PI denotes Proportional Integral, VCO is Voltage-Controlled Oscillator and AWG Arbitrary Waveform Generator. The two plots show the waveforms used to produce a sinc excitation. The phase shifter used is a Mini-Circuits SPHSA-251+ component. Upper panel: intensity waveform produced by AOM 0. Lower panel: phase shift applied to AOM 1. 2.5 V corresponds to a $\pi$ phase shift. (b) The hBEC is highly confined along the vertical direction hence has a broad momentum distribution, much larger than the typical width of the Bragg pulses used here. Thus only a part of the distribution is transferred. The vBEC has a narrow distribution along $z$ and thus acts as a spectroscopic probe of the laser pulse distribution.}
\label{fig:setup}
\end{center}
\end{figure*}
In our experiment, we use a metastable helium Bose-Einstein Condensate (BEC) in two geometries, shown in Fig.~\ref{fig:setup}. 
The vBEC is in an optical dipole trap and is elongated along the $z$ direction hence has a narrow velocity distribution along this axis~\cite{partridge_bose-einstein_2010} -- indeed its width is negligible for what follows. 
We make use of this BEC to perform the spectroscopy measurements. 
The hBEC is in a magnetic trap elongated along $x$ and has a broad velocity distribution along the vertical axis $z$~\cite{robert_bose-einstein_2001}. 
A $\SI{1083}{\nano\metre}$ laser, red-detuned by $\Delta/2\pi=\SI{-800}{\mega\hertz}$ from the 2$^3$S$_1\rightarrow$\,2$^3$P$_0$ transition is split into two beams that intersect at the atomic cloud with a vertical angle of $\alpha = \SI{31}{\degree}$. 
The Bragg velocity is therefore $2\hbar k/m =\SI{49.6}{\milli\metre\per\second}$  along the vertical axis.
With this detuning and the pulse durations used, excitation to the electronically excited state is negligible. 
To generate the modulated Bragg pulses, the power is controlled by an acousto-optic modulator (AOM) common to both beams and locked to a reference signal using a feedback loop with a $\SI{70}{\kilo\hertz}$ bandwidth (Fig.~\ref{fig:setup}). 
This is a simple way to handle the non-linear response of the AOM while also compensating for intensity drifts.%Since the relationship between the laser beam power and the RF signal sent to AOM 0 is non linear and pulse shaping requires a specific intensity waveform, it is necessary to lock the beam power on a voltage setpoint (instead of a ``calibrate and set'' strategy). The servo loop bandwidth is sufficient for the typical waveforms required.
The relationship between the (two-photon) Rabi frequency and the power is calibrated using Rabi oscillations. 
%A potential extension of this setup is to include a feedback loop on the phase shifter.
%\charlie{The calibration is robust as laser power feedback after AOM 0 ensures a good intensity stability. Only a slow drift deviations was observed after an extended duration (typically a few days), due to power fluctuations of the beams after AOMs 1 and 2. This issue can be addressed either by calibrating again the Rabi frequency, or by adding a feedback loop on the power setpoint while measuring the intensity of the Bragg beams after AOMs 1 and 2. The experimental setup does not enhance phase stability.}

To produce a sinc-shaped two-photon Rabi frequency, the laser power at the output of AOM 0 is controlled to be proportional to $|\Omega_{\mathrm{R}}(t)|$ with $\Omega_{\mathrm{R}}(t)$ of Eq.~\ref{eq:sinc}, and a $\pi$ phase shift is added whenever the Rabi frequency changes sign (see Fig. \ref{fig:setup}). The intensity and phase setpoints are sent numerically to an Arbitrary Waveform Generator which is triggered at a specific time after the trap cutoff. The temporal accuracy of both the trigger timing and the generated waveform are better than \SI{1}{\micro\second}. The pulse shapes are limited by the 70\,kHz bandwidth of the power servo loop. 

%%%%%%%%%%%%%%%%%%
% Data
%%%%%%%%%%%%%%%%%%
\section{Experimental results}
\subsection{Pulse shaping results}

After producing the vBEC, we turn off the trap and allow the cloud to expand for 1 ms.
We then apply a velocity independent Raman pulse to transfer the atoms from the $m_J=1$ state to the $m_J=0$ state, rendering the falling cloud insensitive to magnetic field gradients, while the 1\,ms expansion reduces the effect of interactions. 
We then apply the Bragg pulse, including a frequency chirp to compensate the acceleration due to gravity. After diffraction, the atoms fall 46 cm onto a Micro-Channel Plate (MCP) detector~\cite{schellekens_hanbury_2005} which records the arrival times and transverse positions of individual atoms. 
The $\sim\,\SI{300}{\milli\second}$ time of flight is long enough that the detected times and positions correspond to the 3 dimensional velocities of the atoms after the diffraction pulse. 
Due to the narrow velocity distribution ($\SI{3}{\milli\metre\per\second}$),
the cloud is uniformly diffracted into momentum states with an upward Bragg velocity and barely expands along the vertical axis during the time of flight.
The diffracted atoms fall onto the MCP about $\SI{5}{\milli\second}$ after the undiffracted atoms.

\begin{figure}
    \centering
    \includegraphics[width=1\linewidth]{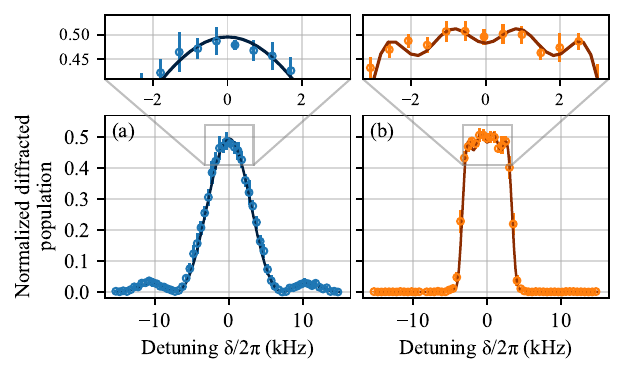}
    \caption{Experimental (dots) and theoretical (solid lines) transfer efficiency of a beam splitter for a square pulse (a) and a sinc pulse (b) where $\SI{1}{\kilo\hertz}$ in detuning corresponds to $\SI{2}{\mm\per\second}$ in velocity. The theoretical expectations  are computed from the Schrödinger equation using the Hamiltonian given in Eq.~\ref{eq:Braggcoupling} without any fit parameter, and integrated over a range of $\SI{1}{\kilo\hertz}$ to account for the experimental binning range. 
    Parameters are $\Omega_{\mathrm{M}}/2\pi=\SI{1.88}{\kilo\hertz}$ and durations of $\SI{133}{\micro\second}$ and $\SI{1}{\milli\second}$ respectively for the square and sinc pulses.}
    \label{fig:splitter}
\end{figure}

\begin{figure}
    \centering
    \includegraphics[width=1\linewidth]{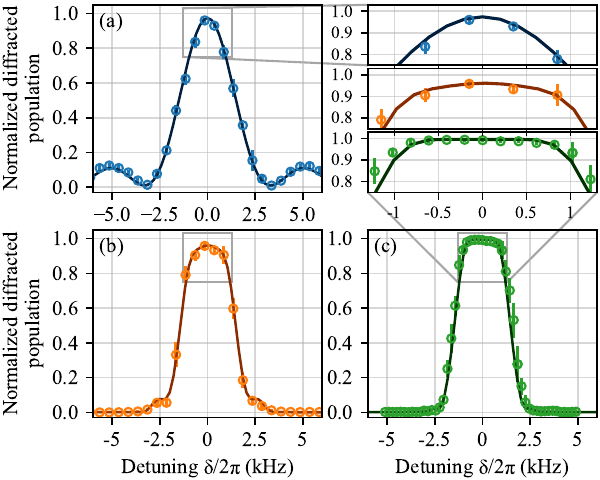}
    \caption{Experimental (dots) and theoretical (solid lines) transfer efficiency of a deflector for a square pulse (a), a sinc pulse (b) and a reburp pulse (c). The theoretical expectations  are integrated over a range of $\SI{1}{\kilo\hertz}$ to account for the experimental binning range. Parameters are $\Omega_{\mathrm{M}}/2\pi=\SI{1.88}{\kilo\hertz}$ and a duration of $\SI{266}{\micro\second}$ for the square pulse,  $\Omega_{\mathrm{M}}/2\pi=\SI{2.05}{\kilo\hertz}$ and a duration of $\SI{1.5}{\milli\second}$ for the sinc pulse, and $\Omega_{\mathrm{M}}/2\pi=\SI{0.57}{\kilo\hertz}$ and a duration of $1.8\,$ms for the reburp pulse. For the reburp pulse, all points in the interval [\SI{-0.5}{\kilo\hertz}, \SI{0.5}{\kilo\hertz}] are better than 98\%.}
    \label{fig:mirror}
\end{figure}
To illustrate the effect of pulse shaping, we scan the frequency difference $\omega_2-\omega_1$, hence the detuning $\delta$, and observe the fraction of diffracted atoms for several pulse shapes. 
The results are shown in Figs.~\ref{fig:splitter} and \ref{fig:mirror} and compared to theoretical expectations, computed without any fit parameter by integrating the Schrödinger equation using the coupling Hamiltonian defined in Eq.~\ref{eq:hsim} for a multi-level model.
%The  atom number and velocity distribution of the atomic source were stable enough to merge the results over several experimental runs without particular correction or post-selection. 
The error bars provided in Figs.~\ref{fig:splitter} and \ref{fig:mirror} account for the typical standard deviation obtained over about 10 runs.

For a $\SI{50}{\percent}$ transfer, which can be thought of as a beam splitter, we compare a square pulse in panel (a) of Fig.~\ref{fig:splitter} with a sinc pulse in panel (b) with the same two-photon Rabi frequency (and therefore the same peak power). Although the fraction of transferred atoms is not small,
%Eq.\ref{eq:FourierTransform} remains approximately valid: 
we observe that the sinc pulse eliminates the side-lobes and leads to an almost square profile. 
The rising and falling slopes of the sinc pulse are 4 times greater than for the square pulse. 
The resonance width, which we define to be the range over which the transferred fraction is close to 1/2 (between 47.5 and 52.5\%), is greater for the sinc pulse by a factor of 1.5.
The results are in very good agreement with the expected theoretical profiles in terms of width, efficiency, and spectral shape. 
Although it was not used to obtain the data in Fig.~\ref{eq:Braggcoupling}, pulse shaping also lends itself easily to apodization which would help to further flatten the spectrum for the sinc pulse. 

We perform the same experiment for a pulse with $\SI{100}{\percent}$ transfer (a momentum deflector). As mentioned in section II, the effectiveness of the sinc-shaped pulses is directly due to the Fourier relationship between the temporal profile of the two-photon Rabi frequency and the momentum response to the pulse. The Fourier relationship is not exact for large population transfers, and so other shapes have been developed, one of which we discuss below.
%Since this idea is less valid for high transfer percentages, additional shapes were designed in other contexts to compensate for the deviations from the Fourier relationship.}
%
We compare a square pulse to a sinc pulse and to a so called \textit{reburp} pulse \footnote{Re-BURP stands for Refocusing Band-Selective Pulse with Uniform Response and Phase} which was identified in the context of NMR~\cite{mcdonald_uses_1991,geen_1991_band-selective} and theoretically studied for Bragg diffraction in Ref.~\cite{luo_contrast_2016}. 
The reburp pulse is defined in terms of a Fourier series as

\begin{equation}
    \Omega_{\mathrm{R}}(t)=\Omega_{\mathrm{M}}\left[A_0+\displaystyle\sum_{n=1}A_n\,\mathrm{cos}(n\Omega_{\mathrm{S}}t)\right]
    \label{eq:reburp}
\end{equation}
for $0\leq t\leq 2\pi/{\Omega_{\mathrm{S}}}$, where $\Omega_{\mathrm{S}}=2A_0\Omega_{\mathrm{M}}$ and the $A_n$ are coefficients up to the 15th order \footnote{The coefficients used are $A_n=[$0.48,$-$1.03, 1.09, $-$1.59, 0.86, $-$0.44, 0.27, $-$0.17, 0.10, $-$0.08, 0.04, $-$0.04, 0.01,$-$0.02, 0.00, $-$0.02$]$. The peak Rabi frequency is 6.24 times $\Omega_\mathrm{M}$.}. 
Like the sinc, this pulse also undergoes sign changes.

The parameters of the three pulses (power and duration) were chosen so as to have the same half width in momentum. 
The results are plotted in Fig.~\ref{fig:mirror}. 
It is observed that for the sinc pulse, the deviations from the Fourier relationship shown in Eq.~\ref{eq:FourierTransform} become significant.
Although not giving a square spectrum, the sinc still reduces the sidelobes and gives a sharper and flatter profile than the square pulse: the slopes at a $\SI{50}{\percent}$ transfer are 1.8 times greater for the sinc pulse, while the resonance width (defined here as the range for which there is at least a $\SI{95}{\percent}$ transfer) increases by a factor of 1.5 compared to the square pulse.
The reburp pulse leads to a momentum deflector for which the resonant momentum range is wider (by a factor of 2 compared to the square pulse), flatter, and sharper (the rising and falling slopes are 2 times greater than the square pulse) than the others.   
We know of no equally effective pulse shapes in the case of $\SI{50}{\percent}$ transfer~\cite{luo_contrast_2016}. 

\subsection{Application: dual coupling}

Pulse shaping also allows one to select two distinct momentum doublets from a distribution. This can be achieved with a single pair of Bragg beams modulated by a cosine function. In the case of a sinc pulse, we have:
\begin{equation}
\Omega_{\mathrm{R}}(t)=\Omega_{\mathrm{M}}\, \mathrm{sinc}[\Omega_{\mathrm{S}}(t-T/2)] \,\mathrm{cos} [\Omega_{\mathrm{D}} t/2]
\label{eq:modulatedpulse}
\end{equation}
From the interaction Hamiltonian given in Eq.~\ref{eq:Braggcoupling}, one can see that a two-photon Rabi-frequency $\Omega_{\mathrm{M}}\,\mathrm{e}^{\mathrm{i}\Omega_{\mathrm{D}} t/2}$ results in an effective detuning which will shift the resonance by $\Omega_{\mathrm{D}}/2$. 
Therefore multiplying any given pulse by a cosine induces a  resonance with two momentum doublets, provided that the duration of the pulse is long enough.
The frequency $\Omega_{\mathrm{D}}$ controls the separation between the selected momentum doublets.

\begin{figure}[tbp]
\begin{center}
\includegraphics[width = \linewidth]{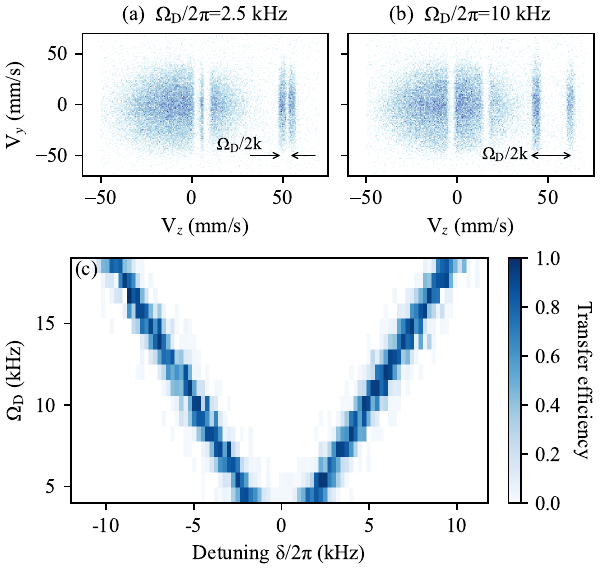}
\caption{Effect of an overall modulation of the diffraction pulse.  Two velocity doublets are selected by the same pulse ($\Omega_{\mathrm{M}}/2\pi=1.5$\,kHz, $T=1.5\,$ms) and the separation is controlled by the modulation frequency $\Omega_{\mathrm{D}}$. (a-b) A  modulation frequency of $\Omega_{\mathrm{D}}/2\pi=2.5\,$kHz leads to a velocity difference of 5\,mm/s in (a) while a modulation frequency of $\Omega_{\mathrm{D}}/2\pi=10\,$kHz leads to a velocity difference of 20\,mm/s in (b). Data is averaged over 50 experimental runs with hBEC. (c) The transfer efficiency of vBEC (color scale) is shown as a function of the detuning. Each slice shows a different modulation frequency.} 
\label{fig:TwoVelocities}
\end{center}
\end{figure}

We illustrate in Fig.~\ref{fig:TwoVelocities} this process experimentally using the hBEC for two values of $\Omega_{\mathrm{D}}$. As expected, there are two resonant velocity classes, separated by $\Delta v=\Omega_{\mathrm{D}}/(2k)$. 
Using the vBEC, we tested this technique over a large range of $\Omega_{\mathrm{D}}$ and confirmed the expected linear variation of the selected velocity class difference by varying the detuning between the two Bragg beams. The frequency difference between the observed resonance peaks as a function of $\Omega_{\mathrm{D}}$ is fitted, and we find a linear relationship with a slope of 
%$1.02\pm0.04$
$\num{1.02(4)}$, which confirms that the modulation frequency $\Omega_{\mathrm{D}}$ indeed controls the resonance difference.\\

\subsection{Differential phase control}

We can also tune the phase imprinted on the atomic wave packets. To this end, a phase parameter $\theta$ can be added to the modulation function:
\begin{equation}
\Omega_{\mathrm{R}}(t)=\Omega_{\mathrm{M}}\, \mathrm{sinc}[\Omega_{\mathrm{S}}(t-T/2)] \,\mathrm{cos} [(\Omega_{\mathrm{D}} t+\theta)/2]
\label{eq:modulatedpulse2}
\end{equation}
where $\theta$ controls the relative phase imprinted between the two selected momentum doublets through its contribution to the phase in Eq.~\ref{eq:rabifreq}. 
The phase imprinted is $ \varphi_L\pm\theta/2$, depending on the considered momentum doublet.

To investigate this phase imprinting effect, we have realized an interferometer using the Bragg pulses. In the following, the procedure will be described in two main steps. First, we will describe the interferometer and the results that were obtained using unmodulated pulses as defined in Eq.~\ref{eq:sinc}. The observation of an interference patterns aims at confirming that a stable phase can be imprinted on the atoms. Second, we will show that the use of modulated pulses like in Eq.~\ref{eq:modulatedpulse2} thereby realizes two parallel interferometers, each involving a different momentum doublet. The objective here is to ensure precise control over the phase difference between these two doublets through the pulse shape parameter $\theta$.

\begin{figure}[tbp]
\vskip 10 pt
\begin{center}
\includegraphics[width=1\linewidth]{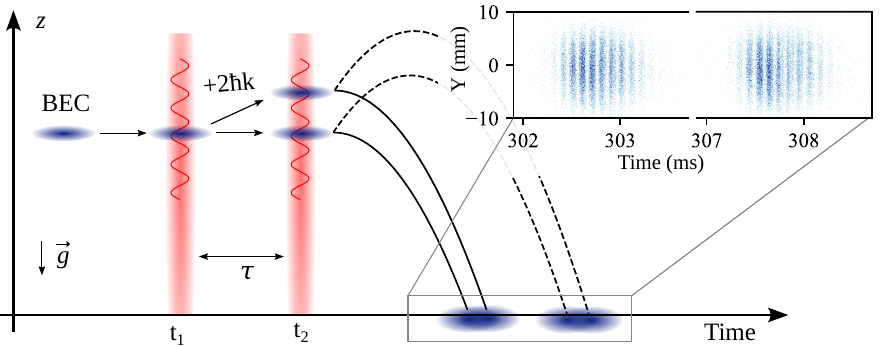}
\caption{Diagram of the interferometer used to test the phase stability of the Bragg pulses.
Two $\pi /2$ pulses create four falling clouds interfering two by two when they overlap at the detector. The fringe period depends on the value of the interferometer time $\tau$. The inset displays fringes observed with the interferometer for $\tau=2\,\mathrm{ms}$, which is much smaller than . The color encodes the density as a function of the arrival time $T$ defined in Eq.~\ref{eq:interferometerPhase} and the horizontal position $Y$. Data is averaged over 25 repetitions: the good contrast confirms the stability of the phase difference for a duration of the order of $\tau$.}
\label{fig:SingleRamsey}
\end{center}
\end{figure}

\paragraph*{Un-modulated pulses:}The interferometer consists of two consecutive beam splitter pulses, as shown in Fig.~\ref{fig:SingleRamsey}.
An hBEC is first split into two parts by a beam splitter sinc pulse similar to that in Fig.~\ref{fig:splitter}.
After a time $\tau$, a second identical pulse is applied and the resulting four clouds fall on the detector. 
Two clouds with the same momentum after the second pulse ($p$ or $p+2\hbar k$) have a spatial separation of $2\hbar k \tau/m$ much smaller than their spatial width, so they overlap. Since they did not acquire the same phase during their fall, the two clouds interfere and  produce fringes while falling on the detector. The interference pattern depends on a phase $\Phi$ given by
\begin{equation}
\Phi=2kg\tau T-\phi_1+\phi_2+\phi_{\mathrm{grav}}
\label{eq:interferometerPhase}
\end{equation}
where $g$ is the acceleration of gravity, $T$ the arrival time at the detector, $\phi_i$ the phase imprinted by each pulse on the atoms ($i=1$ or 2). The constant term $\phi_{\mathrm{grav}}$ corresponds to the relative phase accumulated between the two Bragg pulses. In a gravity field, it depends on $g$ and $\tau$ but not $T$. Propagating the phase of both clouds from their position right after the second pulse to the detector leads to an additional phase difference $2kg\tau T$.

In Fig.~\ref{fig:SingleRamsey} we show the interference patterns corresponding to two overlapping clouds for a wait time of $\tau = 2\,$ms, obtained using two successive sinc beam splitters with $\Omega_{\mathrm{M}}/2\pi=$ 5\,kHz. 
The fringes show high contrast even when averaged over 25 repetitions. 
For each Bragg pulse,  the phase $\phi_i$ imprinted by the pulse is the phase difference $\varphi_L$ between the lasers, so the stability of the fringes confirms that the laser phase difference is stable on a timescale of 2 ms.

\begin{figure}[tbp]
    \centering
    \includegraphics[width=1\linewidth]{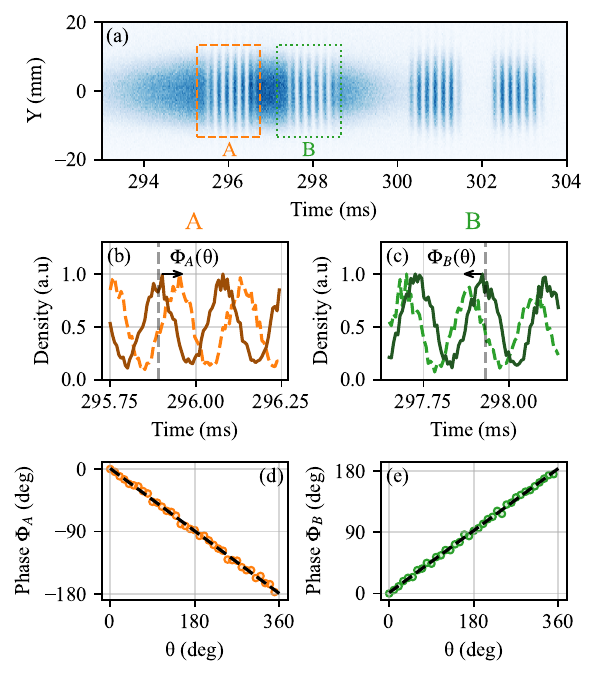}
    \caption{(a) Interference fringes from two parallel interferometers, which are produced by modulated $\pi/2$ pulses as in Eq.~\ref{eq:modulatedpulse}. Data averaged over 350 experimental runs, conducted with a modulated sinc pulse with $\Omega_{\mathrm{M}}/2\pi=5\,$kHz, $\Omega_{\mathrm{D}}/2\pi=10\,$kHz, $\tau=4$\,ms. (b-c) We zoom on the interference region A in (b) and B in (c) to show the fringes for a phase $\theta = 0$ (solid line) and $\theta = \pi/2$ (dashed line). The phase of the interference patterns, defined in Eq. \ref{eq:interferometerPhase}, shifts with $\theta$. (d-e) Phase of the interference pattern as a function of $\theta$, for the interference region $A$ (d) and for the interference region $B$ (e). A linear fit yields slopes of $-0.51(2)$ for $A$ and $+0.50(2)$ for~$B$. The parameters used for these two plots are $\Omega_{\mathrm{M}}/2\pi=1.5\,$kHz, $\Omega_{\mathrm{D}}/2\pi=10\,$kHz, $\tau=1$\,ms.}
    \label{fig:DoubleRamsey}
\end{figure}

The fringes can be shifted at will by adding a voltage to the phase shifter during the second pulse. We observe that the atomic phase $\Phi$ varies linearly with the electronically added phase with slope 1. 
\paragraph*{Modulated pulses:}More importantly, when using pulses modulated by a cosine function as in Eq.~\ref{eq:modulatedpulse2},
we create two parallel interferometers $A$ and $B$ involving different momentum doublets, see Fig.~\ref{fig:DoubleRamsey}. For the first pulse, we use $\theta=0$, leading to $\phi_1^{A}=\phi_1^{B}=\varphi_L$. For the second pulse, we add a phase at the origin $\theta/2$ to the cosine modulation function.
We denote $A$ (resp. $B$) the momentum doublet whose resonance was shifted by $-\Omega_{\mathrm{D}}/2$ (resp. $+\Omega_{\mathrm{D}}/2$). Phases $\mp\,\theta/2$ are therefore imprinted on the two momentum doublets with opposite signs. 
The phases $\phi_2$ in Eq.~\ref{eq:interferometerPhase} are given by:
\begin{equation}
    \phi_2^A =\varphi_L- \theta/2\,  \qquad\text{and}\qquad \phi_2^B =\varphi_L + \theta/2
\end{equation}
where the laser phase difference $\varphi_L$ was shown to remain constant over the time scale of the interferometer. 
Therefore we have:
\begin{equation}
\phi_2^A - \phi_1^A  = -\theta/2\qquad\text{and}\qquad \phi_2^B- \phi_1^B  = \theta/2
\end{equation}
such that the fringes from each resonant doublet are shifted in opposite directions when varying $\theta$, in a way that is independent of the two laser phases.

Panel (a) of Figure~\ref{fig:DoubleRamsey} shows the resulting fringes.  
As expected, we observe four regions exhibiting interference fringes, corresponding to two parallel interferometers. The interference patterns in region $A$ centered at an arrival time of 296 ms and region $B$ centered at 298 ms are respectively shifted by $\mp\Omega_{\mathrm{D}}/2$ relative to the resonance that would be obtained without modulating the Rabi frequency. 
In order to verify that a different phase is indeed imprinted on these two momentum doublets, we fit the interference patterns for different values of $\theta$. 
The phase of the interference pattern is plotted as a function of $\theta$ in panels (d) and (e) of Fig.~\ref{fig:DoubleRamsey} for each of these clouds.  The slopes of the linear fits as a function of $\theta$ are 
$-0.51(2)$ for $A$ and
$+0.50(2)$ for $B$. 
This confirms that the phase at the origin $\theta$ of the pulse shaping modulation function controls the relative phase imprinted between the two selected velocity classes.

The ability to control the relative phase of two parallel beam splitters is of particular importance in an experiment such as that of Ref.~\cite{Dussarrat_two-particle_2017}.

In this experiment, a BEC emits a superposition of atom pairs with opposite momenta \textit{via} a four-wave mixing process~\cite{campbell.2006.parametric,bonneau.2013.tunable}. 
If the population is low, the output state is $\ket{\psi}=\frac{1}{\sqrt{2}}(\ket{p,-p}+\ket{q,-q})$ when restricting the study to two atom pairs $(p,-p)$ and $(q,-q)$. 
This two-particle four-mode state can be used as an input of a Bell interferometer when coupling $p$ and $-q$ on one hand  (momentum doublet $A$), and $q$ and $-p$ on the other hand (momentum doublet $B$). A test of Bell inequality can therefore be realized~\cite{lewis.2015.proposal}, provided that a control parameter can tune the two-particle interference, typically through the phase imprinted on the atoms by Bragg beam splitters. 
This was done in Ref.~\cite{thomas_matter-wave_2022} 
and very recently in Ref.~\cite{athreya2025bellcorrelationsmomentumentangledpairs}. Notably, Ref.~\cite{athreya2025bellcorrelationsmomentumentangledpairs}
% where the control parameter is the sum of the phases imprinted on momentum doublets $A$ and $B$ by a single beam splitter common to both doublets, but the amplitude of the correlation was not high enough to exhibit }
% \victor{nonlocal correlations. 
% Upon resubmitting this work, a preprint reproducing~\cite{thomas_matter-wave_2022} appeared, reporting 
reported a non-local Bell correlations between momentum entangled massive particles .  
However, the \textit{independent} control of the phase of each momentum doublet was not achieved, preventing the authors from demonstrating a violation of the CHSH-Bell inequality \cite{chsh.1969}.

%entanglement or violation of Bell inequality. }

%In the configuration proposed in Ref.~\cite{Dussarrat_two-particle_2017}, the two-particle interference depends on the relative phase of two simultaneously
%applied beam splitters on doublets $A$ and $B$. 
%This phase was fixed in  Ref.~\cite{Dussarrat_two-particle_2017}. With the setup presented in this paper with a dual beam splitter as demonstrated in Fig. 6, this relative phase is well controlled for a given pair of momentum doublets.
%\victor{By controlling the absolute phase of the Bragg laser beam as in Ref.~\cite{athreya2025bellcorrelationsmomentumentangledpairs}, independent control of the phase of each momentum doublets $A$ ad $B$ is achieved, paving the way towards a violation of the CHSH-Bell inequality.}
%
%
%In that experiment, it was necessary to control the relative phase of two simultaneously applied beam splitters. This was done with a single beam splitter and a broad range of velocities.
%The relative phase shift was related to the detuning of each velocity class and was not under the experimentalists control.
%With a dual beam splitter as demonstrated in Fig.~\ref{fig:DoubleRamsey}, this phase is under control. 
%In addition, we have performed simulations of a Bell inequality experiment and shown that not only is the phase controlled but also that the phase remains nearly constant over the velocities inside a given momentum doublet ensuring that all the atoms in the doublet contribute to the Bell signal~\cite{Leprince_thesis_2024}.

In the configuration proposed in Ref.~\cite{Dussarrat_two-particle_2017}, such an inequality requires to control independently  $\phi_A$, the phase of doublet A and $\phi_B$, the one of doublet B. The Bell parameter involved in the CHSH inequality varies then as $\phi_A-\phi_B$ for the above entangled state. 
With the setup presented in this paper with a dual beam splitter
as demonstrated in Fig.~\ref{fig:DoubleRamsey}, this relative phase is well
controlled on contrary to Ref.~\cite{Dussarrat_two-particle_2017} where it was fixed.  Moreover,  
by
controlling the absolute phase of the Bragg laser beam
as in Ref.~\cite{athreya2025bellcorrelationsmomentumentangledpairs}, independent control of $\phi_A$ and $\phi_B$ will be achieved, paving the way
towards a violation of the CHSH-Bell inequality. In addition, we have performed simulations of a Bell inequality
experiment and shown that not only is the phase controlled but also that the phase remains nearly constant
over the velocities inside a given momentum doublet ensuring that all the atoms in the doublet contribute to the
Bell signal~\cite{Leprince_thesis_2024}.

%%%%%%%%%%%%%%%%%%%%%%%%%%%%%%%%%%%%%%

\section{Conclusion}
We have demonstrated precise control over the reflectivity of Bragg diffraction using shaped pulses. Our experimental setup provides access to negative or even complex two-photon Rabi frequencies, thereby enhancing the selectivity and reflectivity characteristics of Bragg transfers. For beam splitters, a sinc pulse produces a square-shaped spectrum, while for deflectors, a reburp pulse yields a more nearly square profile than a sinc pulse. These pulses offer the advantage of being parameter-sparse and easily adaptable to various experimental conditions. 

By modulating a pulse with a cosine function, dual Bragg coupling with resonances with two momentum doublets can be achieved. An interferometry experiment further demonstrates fine control over the phase difference imprinted between each momentum doublet, ensuring that this difference remains, by design, independent of the phases of the lasers used. This is of particular interest when trying to act differently on two momentum classes that are very close spatially.

\section*{Acknowledgments}
We are grateful to Alexandre Dareau, Marc Cheneau and Paul Paquiez for their early contributions to the development of this setup.
The research leading to these results has received funding from QuantERA Grant No. ANR-22-QUA2-000801 (MENTA) and ANR Grant No. 20-CE-47-0001-01 (COSQUA), the LabEx PALM (ANR-10-LABX-0039PALM), Région Ile-de-France in the framework of the DIM SIRTEQ program, and the Quantum Saclay program, supported by the state under France 2030 (reference ANR-21-CMAQ-0002).

\appendix

\bibliography{PulseShaping}

\end{document}